\documentstyle[prb,aps,preprint]{revtex}
\tightenlines 

\begin{document}

\draft

\title{Skyrmions in quantum Hall ferromagnets 
as spin-waves bound to unbalanced magnetic flux quanta} 

\author{J.H. Oaknin, B. Paredes and C. Tejedor}

\address{Departamento de F\'{\i}sica Te\'orica de la Materia
Condensada,
Universidad Aut\'onoma de Madrid, Cantoblanco, 28049, Madrid, Spain.}


\maketitle

\begin{abstract}
A microscopic description of (baby) skyrmions in quantum Hall 
ferromagnets is derived from a scattering theory of collective
(neutral) spin modes by a bare quasiparticle. 
We start by mapping the low lying spectrum of spin-waves in the uniform
ferromagnet onto that of free moving spin excitons, and then we study 
their scattering by the defect of charge. 
In the presence of this disturbance, the local spin stiffness varies in space, 
and we translate it into an inhomogeneous metric in the  
the Hilbert space supporting the excitons.
An attractive potential is then required to preserve the
symmetry under global spin rotations, and it traps the excitons around the
charged defect. The quasiparticle now carries a spin texture. 
Textures containing more than one exciton are described within a 
mean-field theory, the interaction among the excitons being taken into account 
through a new renormalization of the metric. 
The number of excitons actually bound depends on the Zeeman coupling, that
plays the same role as a chemical potential. 
For small Zeeman energies, the defect binds many excitons which 
condensate. As the bound excitons have a unit of angular momentum, provided
by the quantum of magnetic flux left unbalanced by the defect of charge,
the resulting texture turn out to be 
a topological excitation of charge 1. Its energy 
is that given by the non-linear sigma model for the ground state in this
topological sector, i.e. the texture is a skyrmion. 
\end{abstract}

\pacs{PACS numbers: 73.40.Hm}


\section{Introduction}

Low energy excitations of two dimensional systems in the quantum Hall regime  
reveal a rich variety of new physical phenomena. Specially striking 
are the cases of filling factors which imply a ground state which is 
a quantum Hall ferromagnet (QHF) in the lowest Landau level (LLL), 
because their 
charged quasiparticles may carry non-trivial spin textures. In the limit 
of low Zeeman coupling, these textures have been identified 
as skyrmions with real and topological charge equal to one, 
being successfully described by means of a nonlinear $\sigma -$model 
(NL$\sigma$M).\cite{Sondhi}  
For realistic Zeeman energies, the quasiparticles may be regarded as distorted 
(baby) skyrmions with the same topological winding number but a finite 
size determined by the competition between the Zeeman coupling and the electronic 
interaction. Classical field theories fail to give an accurate description 
of these localized spin textures for which quantum fluctuations are
important. Various microscopic approaches have been 
used to attack this task: (i) mean field descriptions of the 
Hartree-Fock type\cite{Fertig} (ii) different kinds of variational 
schemes\cite{Palacios,Abolfath} and (iii) microscopic wave-functions, with well 
defined quantum numbers, obtained from small systems.\cite{Oaknin}  

In this paper we face the problem of baby-skyrmions in actual QHF from a 
completely different point of view. 
We present a description of these
textures derived from a scattering theory of collective (neutral)
spin modes by a bare charged defect. By bare charged defect we mean the 
quasiparticle in the case of  
Zeeman coupling much larger than the electronic
interaction. 
Our scheme starts with long wavelength spin-waves in
the uniform QHF, where they have been shown to behave as free moving 
electron-hole {\it eh} pairs (excitons) of vanishing electric dipolar moment. 
These excitons exhaust the low lying spectrum of neutral excitations of 
the QHF. Our aim is to show that these modes may destabilize the 
bare quasiparticle by becoming bound to it and raise a charged spin texture. 
We have developed an effective Hamiltonian to describe this scattering event. 
The actual number of bound excitons, {\it i.e.} of spin flips, in the 
lowest energy texture is a matter of the Zeeman coupling, that in our picture 
plays the role of a chemical potential. 
To consider skyrmions as spin-waves bound to defects has been 
sometimes suggested but never developed.\cite{Abolfath,Palacios2} 

Before going into details, let us summarize the main ideas of our work.
As mentioned above, the low lying spectrum of a uniform QHF, 
in the strong magnetic field limit, can be mapped to 
that of a gas of free {\it eh} pair of vanishing electrical dipolar
moment and well defined linear momentum.\cite{Kallin} We assume that this  
mapping is still possible in the presence of 
the bare charged defect, although the excitons are not free any more.  
They interact with the disturbance as well as among themselves.
The effect of the disturbance is threefold: 

i) As the quasiparticle is identical to one of the components of the 
exciton (the spin up hole or the spin down electron), Pauli principle does
not allow them to share spatial position. In other words, the exciton moves
in a non-uniform space because its position is less probable near the defect
than far from it. We will show that this leads to an inhomogeneous metric in the 
Hilbert space supporting the wave function of the exciton.

ii) The extra charge of $\mp \nu $ electrons leaves unbalanced exactly one 
quantum of magnetic flux. It introduces an Aharonov-Bohm phase that gives a
unit($\pm 1$) of angular momentum to the excitons in their ground state. 
Mapped back to the language of spin textures it means vorticity with 
winding number equal to $\pm 1$.

iii) Finally, the sharp localization of the spin of the bare quasiparticle, 
facing that of the particles in the QHF (either electrons or holes),
is a  waste of energy, since a smoother alignment of spins would allow
to gain exchange energy.
It produces an attractive short-range potential that binds the excitons. 

Using simple arguments, we derive in 
this work explicit expressions for both the new metric and the potential. 
This kind of treatment, is enough to find the lowest lying bound state 
of a unique exciton. Comparison of this state with the one obtained by 
numerical calculations firmly supports our framework. 

The next step required for a complete description of charged textures is to  
let the defect bind more than one exciton,  
what implies to take into 
account the interaction among them. 
For that purpose we use a mean field approximation for the excitons. 
This approach should be very accurate for textures containing a large
number of excitons, in which case we should recover a description for
skyrmions of quasi infinite size. We will check that effectively this is our
result. As well, it follows from our analysis that it also works quite well even 
for textures with very few excitons. 

The paper is organized as follows: In section II we describe the scattering
of one exciton by a charged defect. 
Since the method does not require a microscopic wave function of the QHF, 
it is directly applicable to all the cases with filling factor 
$\nu =1/(2p+1)$ with $p$ integer. For the sake of simplicity the main body of the 
paper is devoted to $\nu =1$, but a discussion of the case $\nu =1/3$ is also 
presented at the end of that section II. In order to study a spin
texture with a larger size, section III presents our mean field approach 
for the binding of an increasing number of excitons. 
A brief summary is presented in section IV. 

\section{One spin-wave bound to a charged defect} 
Let us start by describing a spin-wave in a uniform QHF in terms of 
a single exciton within the formalism later required to study its
scattering by
a charged defect. We consider the magnetic field high enough to make
the LLL approximation. Since the system is invariant under spin rotations 
around a unitary vector ${\bf u}_B$ in the direction of the magnetic field, 
the Zeeman coupling {\em \~{g}} gives only an energy shift. Therefore, 
{\em \~{g}} is for the moment taken as zero and included a posteriori. 
As the kinetic energy is quenched in the LLL approximation, the only
contribution to the Hamiltonian comes from the interaction between
electrons.

\subsection{Single exciton in a uniform ferromagnet} 
Since all the polarized excitations of the QHF have a finite gap, its 
lowest lying spectrum is exhausted by the collective spin modes 
associated to the spontaneous breaking of the rotation symmetry in spin space
(spin-waves). These long wavelength textures are made up of non interacting
single spin excitons 
in which one electron flips its
spin to down leaving behind a hole in the filled spin up level. The most general
form of this kind of excitations is 
\begin{eqnarray}
O^\dagger _\phi |F\rangle = \int d{\bf r}d{\bf r'} 
\phi ({\bf r},{\bf r'}) \Psi ^\dagger _\downarrow ({\bf r'})
\Psi _\uparrow ({\bf r})|F\rangle 
\label{oper}
\end{eqnarray}
where $|F\rangle $ is the fully polarized ground state and $ \Psi ^\dagger
_\downarrow $ and $\Psi _\uparrow $ are creation operators of the electron
and the hole, respectively, projected onto the LLL.
The operator $O^\dagger
_\phi $ creates a {\it eh} pair in the positions ${\bf r},{\bf r'}$ 
with a probability given by $|\phi ({\bf r},{\bf r'})|^2$.
The function $\phi ({\bf r},{\bf r'}) $ characterizes completely the
operator $O^\dagger _\phi $, and can be interpreted as the wave function of the 
exciton. As the {\it eh} pair is neutral and $|F\rangle$ is translationally
invariant, $\phi ({\bf r},{\bf r'}) $ may be classified by a conserved wave
vector $\bf k$ that plays the role of the total linear momentum of the
exciton. Within the LLL approximation, the fully interacting Hamiltonian
can be exactly diagonalized to yield:\cite{Kallin}
\begin{eqnarray}
\phi _{\bf k}({\bf r},{\bf r'}) =\frac{1}{2\pi} e^{i{\bf k R}} 
e^{iX\Delta y/l_B^2}{\cal G}(\Delta {\bf r}-l_B^2{\bf k}\times {\bf u}_B) 
\label{funexc}
\end{eqnarray}
${\bf R}=({\bf r}+{\bf r'})/2$ is the center of mass coordinate with
components $(X,Y)$, $\Delta{\bf r}={\bf r}-{\bf r'}$ is the relative coordinate with
components $(\Delta x,\Delta y)$ and $l_B$ is the magnetic length. 
The function multiplying the plane wave is just         
the representation of the delta function in the LLL 
approximation. It is, except for a gauge dependent factor, a gaussian whose 
width is the magnetic length. This widening expresses that in the LLL
the coordinates $x$ and $y$ are conjugated operators, and hence liable to 
uncertainty. 
What Eq. (\ref{funexc}) tells us is that the electron and the hole move parallel
to one another with a constant velocity perpendicular to their electric
dipolar moment, given by $\Delta {\bf r}=l_B^2 {\bf k}\times {\bf u}_B$. 

Our interest is on the long wavelength limit 
$k<<l_B^{-1}$. In this case the distance between the two opposite
charges vanishes, and the corresponding eigenstates read simply 
\begin{eqnarray}
\lim _{kl_B \rightarrow 0} \phi _{\bf k}({\bf r},{\bf r'}) =
\frac{1}{2\pi}e^{i{\bf k R}} \delta (\Delta {\bf r}).
\end{eqnarray}
Now, Eq. (\ref{oper}) becomes:
\begin{eqnarray}
O^\dagger _\phi |F\rangle = \int d{\bf r}
\phi ({\bf r}) \Psi ^\dagger _\downarrow ({\bf r})
\Psi _\uparrow ({\bf r})|F\rangle 
\label{operator}
\end{eqnarray}
with a wave function for the exciton simply given by 
\begin{eqnarray}
\phi ({\bf r}) =\frac{e^{i{\bf k r}}}{\sqrt {2\pi }}.
\end{eqnarray}
It is worth commenting that the normalization is a result of the fact that
\begin{eqnarray}
\langle F|O_\phi O^\dagger _\phi |F\rangle =1 \Longrightarrow 
\int d{\bf r} |\phi ({\bf r})|^2=1.
\label{norm0}
\end{eqnarray}
The energy of these long wavelength excitons results to be 
quadratic on the linear momentum (wave vector),
$\varepsilon _1({\bf k})=4\pi\rho_s k^2$,  
where $\rho _s$ is the spin stiffness of the QHF.\cite{Kallin} 
Since any excitation $O^\dagger _\phi |F\rangle $ is totally characterized by
a function $\phi ({\bf r})$ and energy $\varepsilon _1$, it follows\cite{Oaknin2} 
that the many-body problem we are considering, can be mapped onto the   
single particle Hamiltonian that governs the dynamics of 
a neutral free particle with an effective mass $m^ \ast =(8\pi \rho _s)^{-1}$ 
\begin{eqnarray}
{\cal H}O^\dagger _\phi |F\rangle= \varepsilon O^\dagger _\phi |F\rangle 
\Longleftrightarrow -4\pi \rho _s\nabla ^2 \phi= \varepsilon \phi 
\label{hamilt1}
\end{eqnarray}
${\cal H}$ being the fully interacting many-body Hamiltonian. 

\subsection{Scattering of a single exciton by a defect of charge $\pm 1$}
Our aim is to describe (baby)skyrmions as bound states of spin-waves 
to a charged defect. Hereafter, we take a hole in the ferromagnetic 
ground state as the bare defect. So, we will obtain an 
antiskyrmion, while a skyrmion should be produced by taking as a bare 
defect an electron with opposite spin to that of the electrons in the 
ferromagnetic ground state. Within the framework presented in the previous 
subsection, the problem reduces to study the dynamics of the otherwise 
free moving exciton $\phi _1 ({\bf r})$ in the presence of the bare 
quasiparticle. Physically we expect the defect to influence the exciton 
dynamics in two different ways: 

{\it First}, as the quasihole (quasielectron) 
is identical to one of the components of the exciton (the
spin up hole or the spin down electron respectively), Pauli principle 
forbids them to come together. In other words, the exciton moves
in a non-uniform space because its position is less probable near the defect 
than far from it. As well, the extra charge unbalances the  
commensurability between the number of particles and quanta of magnetic flux.
The unpaired quantum of flux introduces an Aharonov-Bohm phase. 

{\it Second}, the spin up quasihole (spin down quasielectron) 
has no exchange interaction with spin down holes 
(spin up electrons) in the filled (either with holes or electrons) level.
In order to lower the energy, it is preferable to have a smoother spin field
which allows to gain exchange energy. Since the exciton is described by a
spin-flip operator, it feels the exchange field as an attractive effective 
potential in the region of the defect. 

Let us follow these physical ideas to derive an  
effective Hamiltonian
describing
the dynamics of a single exciton, $\phi _1 ({\bf r})$,
in the presence of a bare quasiparticle.
To preserve the symmetry of the problem we use symmetric gauge centered in 
the position of the defect. Single particle wave functions are 
\begin{eqnarray}
\varphi _m(z)= \frac{1}{\sqrt{2^{m+1}\pi m!}}z^me^{-|z|^2/4l_B^2}
\label{varphi}
\end{eqnarray}
where $z=x+iy$, so that hereafter we replace ${\bf r}$ by $z$. 
The extra hole occupies the single particle state with $m=0$
and we denote the state containing the bare quasiparticle 
by $|^0F\rangle $. What is now the Hilbert space of
functions $\phi _1 (z)$ for the exciton? As a consequence of the fact that
$|^0F\rangle $ is not translationally invariant, the normalization condition
is not given by Eq. (\ref{norm0}) any more. Some straightforward algebra
allows to obtain that 
\begin{eqnarray}
\langle ^0F|O_\phi O^\dagger _\phi|^0F\rangle =1 \Longrightarrow 
\int dz|\phi _1 (z)|^2 |\mu _1 (z)|^2 =1
\end{eqnarray}
where the new metric for the Hilbert space is given by 
\begin{eqnarray}
|\mu _1 (z)|^2=1-2\pi l_B^2|\varphi _0(z)|^2
\end{eqnarray}
In other words, there is change in the scalar product due to the presence of
the quasihole occupying $\varphi _0(z)$. This can be understood with the
following physical arguments. For an exciton in a uniform QHF, the
probability amplitude of being at a position $z$ is, of course, given by
$|\phi _1 (z)|^2$. In the presence of a flux quantum, the {\it eh} pair
can not move into the region around the defect. In this case, the probability
of finding the exciton in a position $z$ is $|\phi _1 (z)|^2$ multiplied
by the probability $1-2\pi l_B^2|\varphi _0 (z)|^2 $ of not having 
the defect at the same position. 
This is precisely the new metric $|\mu _1 (z)|^2$ we have derived. 

The function $|\mu _1 (z)|^2$ tends to one for large distances so that
the usual metric is recovered far from the defect. On the contrary, $|\mu _1
(z)|^2$ goes to zero as $r^2$ near the origin, where the defect is
located. As a crucial result of this behavior around the origin, the new
Hilbert space $L^2(\mu _1 )$ of exciton wave functions includes new functions
that were not square integrable with the usual homogeneous metric (\ref{norm0}). In
particular, the Hilbert space is expanded with an exciton function that
diverging at the origin as $r^{-1}$ has now a finite norm. This new
open possibility produces interesting physical results.  

The complex function $\mu _1 (z)$, whose square modulus gives the new metric,
has been obtained hitherto up to a 
phase factor $e^{if(z)}$. To be consistent with the fact that the hole is 
linked to an unbalanced quantum of magnetic flux, we chose this factor to be
the Aharonov-Bohm phase $e^{i\theta }$, so that  
\begin{eqnarray}
\mu _1 (z)= e^{i \theta } \sqrt {1-2\pi l_B^2|\varphi _0 (z)|^2} . 
\label{measure1}
\end{eqnarray}
An obvious implication of the new metric is to alter the effective Hamiltonian 
describing the exciton dynamics. As this Hamiltonian must be hermitian 
with respect to the new metric, the eigenvalue equation must take the form:
\begin{eqnarray}
H\mu _1 \phi _1=\varepsilon _1 \mu _1 \phi _1
\end{eqnarray}
where $H$ is Hermitian with respect to the usual homogeneous metric. 
With this, Eq. (\ref{hamilt1}) would read 
$-4\pi \rho _s\nabla ^2 \mu _1 \phi _1=\varepsilon _1\mu _1 \phi _1$
and the problem would be formally identical to that of the uniform QHF. 
However, we have overlooked an important contribution. In its present  
form, the Hamiltonian does not preserve the
symmetry of being invariant under spin rotations in the spin space. 
We should remember that in our problem the Zeeman and the electronic
interaction are separable, and by now we are only taking into account 
the latest one. Making a rigid rotation of
the spin must then cost zero energy. Such a rigid rotation is generated by an
operator of the form (\ref{operator}) with $\phi _1 (z)$ replaced by a
constant (i.e. the spin lowering operator). Hence, $\phi _1 (z)
\equiv constant$ must be 
a zero energy eigenstate of the single particle effective Hamiltonian. This
condition was guaranteed in the homogeneous metric case, as $\nabla ^2$ 
gives zero when acting on a constant.
However, it is no longer true  
in our new inhomogeneous space, and the effective Hamiltonian needs a position   
dependent potential correcting the contribution coming from $\nabla ^2$ 
\begin{eqnarray}
V(z)=4\pi \rho _s \frac{\nabla ^2\mu _1 (z)}{\mu _1 (z)} .
\label{potential}
\end{eqnarray}
This is an attractive central potential actually capable of
binding the excitons. It increases monotonously
from the center and it is concentrated in a few magnetic lengths. 
This is precisely the potential that on physical grounds we expected to
describe the overcost of exchange energy due to having a bare quasiparticle. 

Therefore, the single particle eigenvalue equation takes the form:
\begin{eqnarray}
4\pi \rho _s \left[ -\nabla ^2 +\frac{\nabla ^2\mu _1 (z)}{\mu _1 (z)}
\right] \mu _1 \phi _1 = \varepsilon _1 \mu _1 \phi _1 . 
\label{effectH}
\end{eqnarray}
This equation for the exciton in the presence of the defect is the
first important result of this paper. For the renormalized 
function $\mu _1 \phi _1$, Eq. (\ref{effectH}) is the usual one for a 
particle in a potential given by Eq. (\ref{potential}). 
The exact matching between one extra quantum of flux and one lacking
charge in the function $\mu _1 (z)$ leads to a finite value of 
(\ref{potential}) at the origin. 
As well, this value turns out to be $V(0)=-4\pi
\rho _s$, which is precisely the energy (measured with respect to
$|^0F\rangle $) of the classical infinite-sized 
skyrmion predicted by the NL$\sigma $M.  
In our description, this energy sets a lower bound to the energy of the 
single exciton described by Eq. (\ref{effectH}) as it should be demanded in
the sake of consistency. 

It is worth commenting that the details of the electron-electron interaction are
contained in $\rho _s$. For instance, for a contact interaction between the
electrons, the spin stiffness is zero and there is neither localizing potential
nor dispersion relation.
(\ref{potential}) Therefore, no excitons are bound to the defect for a 
contact interaction between the electrons. This
is in agreement with the exact result stating that for a contact interaction
between the electrons, the spin textures have zero energy.\cite{MacDonald}

The shape of the potential (\ref{potential}) is shown in Fig. \ref{fig1}.
It is pretty close to $-1/cosh^2(r)$
so that one could approximate the lowest lying 
solution of Eq. (\ref{effectH}) by
polynomial expansions, but the consecutive trapping of more than one exciton,
that we will discuss latter, requires numerical procedures, so that it is
better to start already solving Eq. (\ref{effectH}) also by numerical
methods. In any case, without any numerical calculation, it is possible 
to draw the main properties of the lowest lying eigenstate of (\ref{effectH}): 

\begin{itemize}

\item {\em Energy}: It is a bound state with negative energy

\item {\em Angular momentum}: The product of functions $\mu _1 \phi _1$ 
does not have any angular dependence. 
It implies that the exciton wave function takes the opposite dependence on
$\theta $ to that of $\mu _1 $, i.e. 
\begin{eqnarray}
\phi _1(z)=f(|z|)e^{-i \theta }.  
\end{eqnarray}
The phase factor $e^{-i \theta }$ gives a winding number -1 to the spin
texture $O^\dagger _{\phi }|^0F\rangle $. 

\item {\em Behavior at the origin}: near the origin, the product $\mu _1 \phi
_1$ tends to a finite constant value. As the metric vanishes as $|z|$ in this
region, the limit behavior of the exciton wave function at the origin is 
\begin{eqnarray}
\lim _ {|z|<l_B} \phi _1 (z)= \frac{e^{-i \theta}}{|z|}.
\end{eqnarray}
We must stress that in spite of the divergence at the origin, $\phi_1$ has
finite norm with the new metric $\mu _1 $. 

\item {\em Behavior at infinity}: For distances much larger than $l_B$, the
solution $\mu _1 \phi _1$ decays exponentially to zero. As $\mu _1 $ tends to 1,
the exciton wave function $\phi _1(z)$ also decays exponentially. 

\end{itemize} 

All the above discussed characteristics bring to a very important
conclusion:{\em the bare quasiparticle is able to bind a spin exciton, and 
raise in this way a charged spin texture. In its ground state the exciton has a
unit of angular momentum that tries to cancel the quantum of flux 
left unbalanced by the lack of charge}. As we will discuss later, 
for small Zeeman energies, the defect can bind many of such excitons
which  condensate. Order in the magnetization is developed, the angular dependence
of the in-plane component of the order parameter 
being equal to the angular momentum of the excitons.
In other words, a spin texture with unit topological charge appears. 

In order to check the quality of our description we can start by discussing
the behavior around the origin. The radial dependence 
$1/|z|$ here obtained, 
is precisely the one coming out in exact diagonalizations for small quantum 
dots\cite{Oaknin} 
where the finite size of the droplet 
cuts off the exponential decay of the exciton wave
function, but leaves its core perfectly preserved. 
Even more important, this slow decay describes as well the long distance
behavior of those textures made up of many excitons. 
\cite{Oaknin,MacDonald} The exponential decay
keeps the exciton localized \cite{Palacios,Abolfath}, and the size
of the texture finite. We will see in the next section that, 
as more excitons are bound, their mutual repulsion makes them to spread, 
and the exponential decay of their wave function 
becomes smoother. Then, for textures made up of many excitons,
the wave function of each of them is simply $1/|z|$. 

The energy of the texture created by one exciton is a very interesting 
magnitude because it determines the critical $g$-factor $g_{cr}$ 
for the existence of baby skyrmions. For $g>g_{cr}$, the positive Zeeman 
energy is so high that the exciton becomes unbound and no spin
texture can be formed. For this binding energy we obtain   
$-0.311\times 4 \pi \rho _s$, that would imply 
a $g_{cr}=7.9/\sqrt{B}$ T$^{1/2} $ for the existence of a
baby-skyrmion in a GaAs quantum well at $\nu =1$. This energy is far below
the value $-0.17\times 4 \pi \rho _s$ previously calculated by using 
a variational approximation for the microscopic wave function of the 
charged spin texture.\cite{Palacios,Abolfath}    

Our scheme can be directly applied to a QHF corresponding to a filling
factor $\nu=1/3$, just by changing the energy scale given by the spin
stiffness that now is 27 times smaller than in the case of filling factor
1.\cite{Moon} This implies a critical $g$ factor $g_{cr}=0.29/\sqrt{B}$
T$^{1/2} $ for the existence of a baby-skyrmion in a GaAs quantum well 
at $\nu =1/3$. For a field of a few Teslas, this value is below the
$g$-factor of GaAs. This would explain why skyrmions at $\nu =1/3$ are observed
only if high pressure is applied reducing the $g$-factor,\cite{Leadley} 
while they do not appear in the normal case.\cite{Leadley,Khandelwal} 

\subsection{Scattering of a single exciton by a defect of charge $\pm Z$}

Our model is easily generalized to the case in which $Z>1$ flux
quanta are added or removed from the system in order to create a defect with
higher real and topological charge. In this case the metric is given by  
\begin{eqnarray}
\mu _1 ^{(Z)}(z)=  e^{\pm i Z \theta } \sqrt {1-2\pi l_B^2 \sum _{m=0}^{Z-1} 
|\varphi _m (z)|^2} . 
\end{eqnarray}
$Z$ quasiholes or quasi electrons are occupying $\varphi _m (z)$
states (with $m=0,...,Z-1$) due to the added or removed flux quanta. Now the
metric vanishes as $r^{2Z}$ near the origin, so that an exciton wave function
diverging at the origin as $r^{-Z}$ has finite norm. The potential $V(z)$ 
has the same behavior described for $Z=1$. In particular, its minimum
is equal to the energy of the skyrmion with topological charge $Z$ 
predicted by the NL$\sigma $M, i.e., $-Z4\pi \rho _s$. Once again the
product of functions $\mu _1 ^{(Z)}(z) \phi_1^Z (z)$ in the lowest energy 
solution of Eq. (\ref{effectH}) has no angular dependence, so that 
the exciton wave function behaves as $e^{\mp i Z \theta }$. In other words, 
we find that when a spin-wave becomes bound to a 
defect of physical charge $Z$, it raises 
a spin texture $O^\dagger _{\phi ^Z}|^{Z-1}F\rangle $ with winding number $Z$. 

As an example, we have computed the binding energy of one exciton in a 
charged defect with $Z=2$. We get an energy of $-1.13 \times 4\pi \rho _s$ 
which is much smaller than the result obtained from a Hartree-Fock 
calculation.\cite{Lilliehook}. This energy is important to decide if  
a skyrmion with charge $Z=2$ is cheaper or not 
than two skyrmions of charge $Z=1$ separated a finite distance.  

\section{Several spin-waves bound to a charged defect} 
In the previous section we have focused our attention on spin 
textures made up with only one electron flipping its spin.
Since the Zeeman contribution to the energy is separable from the 
contribution due to the interaction between electrons, one can add 
it {\em a posteriori} so that, hitherto, we have not included it.
The Zeeman energy is proportional to the number of electrons
that have flipped their spin to raise the texture. 
As rotations 
in spin space around the direction of the magnetic field are 
symmetry operations, this is a well defined 
quantum number for the textures. In our language, in which we count each spin
flip as an exciton, the Zeeman coupling constant just plays the role
of a chemical potential. By tuning the $g$-factor, we may change the 
number of excitons actually
present in the lowest lying texture. In this section we are going to study
those textures made up of more than one bound exciton. 
Once again, the Zeeman contribution is added at the end of the process.

We will approach this problem within a mean field theory for the excitons.
Our task then reduces to generalize the scattering formalism we have
already developed, {\it i.e.} Eq. (\ref{effectH}),
in order to describe the dynamics of one single exciton in the 
presence of both the unbalanced flux quantum and a background of the 
remaining excitons. We want to stress that the singled out exciton is identical to
those in the background, and so we  will be led to solve this problem 
self-consistently. It is also worthy to comment that if we
apply the same ideas to the spin-waves in the uniform QHF, 
we obtain that they do not interact because they are completely 
delocalized. This is not the case, nevertheless, for the excitons bound to the
quasiparticle. 
Before going into formal matters, let us 
analyze what we should expect to be 
the effect of a background of excitons in the dynamics of another one coming
into the region of the defect. 

When an exciton gets bound, spin up hole states around
the charge defect begin to be filled up.  
A new exciton moving in this texture will find a density of occupied 
(unaccessible) hole states larger than it would find if there were only a     
bare defect. Therefore, we expect a background of already bound excitons 
to change the effective metric felt by a new coming one   
in order to account for the new unaccessible states. Moreover,  
as some exchange
energy has already been gained by trapping the background excitons, 
we also should expect a less attractive effective potential and a weaker 
energy needed to keep a single exciton tied in the spin texture.     
In other words, we expect 
the background of bound excitons to screen the bare defect. 
As more excitons
get bound, 
the effective screening spreads over a larger region, and so does the area  
where the excitons are constrained to stay.
Eventually, when the flux quantum is almost
completely screened, the energy required to tie or drop a single exciton 
tends to vanish, and the system develops order in the magnetization.
The resulting textures are skyrmions of quasi infinite size, as the ones 
described by a NL$\sigma $M.  

Let us formally derive now the one particle 
effective Hamiltonian for an exciton in the
presence of both a flux quantum and a set of $K-1$ previously bound excitons. 
The first thing to do is to deduce the change in the metric due to the
binding of more and more excitons. At the $K$-th step, the mean field state 
is built up as $(O^\dagger _\phi )^K|^0F\rangle=
O^\dagger _\phi \left[(O^\dagger _\phi )^{(K-1)}|^0F\rangle \right] $.
Note that all the $K$ excitons are in the same state $\phi _K (z)$. The only 
reason for writing the $K$-th separately is to make explicit our aim
of studying the dynamic of a single one of them in the presence of the
remaining ones. 

Following our own steps in the previous section, let us evaluate the
normalization condition for the exciton wave function $\phi _K (z)$
\begin{eqnarray}
1= \frac{1}{K}\frac{\langle ^0F|(O_\phi) ^{(K-1)}O_\phi O^\dagger _\phi 
(O^\dagger _\phi )^{(K-1)}|^0F\rangle }
{ \langle ^0F|(O_\phi) ^{(K-1)} (O^\dagger _\phi )^{(K-1)}|^0F\rangle }
\label{renorm}
\end{eqnarray} 
where the factor $1/K$ appears due to the different possibilities of
singularizing one exciton among the $K$ electrons and holes which are present. 
If the operators 
$\Psi ^\dagger _\downarrow ({\bf r}) \Psi _\uparrow ({\bf r})$ 
and 
$\Psi ^\dagger _\uparrow ({\bf r^{\prime}})
\Psi _\downarrow ({\bf r^{\prime}})$ 
satisfied perfectly 
bosonic commutation relations, Eq.(\ref{renorm}) would imply again the 
condition for normalization of a unique exciton  
\begin{eqnarray}
\int dz |\phi _K (z)|^2 |\mu _1 (z)|^2=1
\end{eqnarray}
with $\mu _1 (z)$ given by Eq. (\ref{measure1}). In fact this is the
case for few spin waves excited on the uniform QHF, and they 
behave as independent bosons. However,  
in the presence of the defect of charge it is not so anymore.  
What we get instead from Eq.(\ref{renorm}) is 
\begin{eqnarray}
\int dz |\phi _K (z)|^2 |\mu _K (z)|^2=1
\end{eqnarray}
where the new renormalized metric is 
\begin{eqnarray}
|\mu _K (z)|^2=|\mu _1 (z)|^2-|\Delta \mu _K (z)|^2 
\label{measureK}
\end{eqnarray}
with the variation of the metric taking the form  
\begin{eqnarray}
|\Delta \mu _K (z)|^2=(K-1) \sum _{m>0} \alpha _m^{(K)}|\varphi _m(z)|^2 .
\label{Deltamu}
\end{eqnarray}
In this expression, $\varphi _m(z)$ are the single-electron states
given by Eq. (\ref{varphi}) and $
\alpha _m^{(K)}$ satisfy 
\begin{eqnarray}
\sum _{m>0} \alpha _m ^{(K)}=1
\label{suma} 
\end{eqnarray}
for any $K$. For each step $K$, $\alpha _m^{(K)}$ is a function 
(given in the Appendix up to $K=4$) of the wave function of 
the exciton through the integrals 
$\int dz|\phi _{K} (z)|^2 |\varphi _{m} (z)|^2$.
Apart from the more or less complicated
expression of these coefficients, Eq. (\ref{measureK}) simply states that the
initial metric $\mu _1 (z)$ is corrected by the non zero occupation of
states $\varphi _m(z)$ with $m>0$ due to the $K-1$ previously bound excitons. 
The condition (\ref{suma}) guarantees that the number $\int dz |\Delta \mu
_K (z)|^2$ of new unaccessible hole states equals the number $K-1$ of previously
bound excitons. Therefore, the coefficients $\alpha _m ^{(K)}$ 
play the role of effective occupations of states $\varphi _m(z)$ in the texture.

It must be pointed out that, as the coefficients $\alpha _m^{(K)}$ 
involve a dependence on the excitonic wave function $\phi _K (z)$, 
$|\mu _K (z)|^2$ is effectively a new renormalized 
metric only for the very same wave function with which we have built it.
Nevertheless, at first order, we may accept that it is also valid for 
all those other wave functions that are close in energy to the former one. 
If the constant function, $\phi(z) \equiv constant$, could be included 
among these, we would just need to repeat the symmetry reasoning of the previous
section to obtain a new effective exchange potential, and so the generalization
of Eq. (\ref{effectH}) we are looking for: 
\begin{eqnarray}
H_K \mu _K \phi _K =
4\pi \rho _s \left[ -\nabla ^2 +\frac{\nabla ^2\mu _K (z)}{\mu _K (z)}
\right] \mu _K \phi _K = \varepsilon _K \mu _K \phi _K . 
\end{eqnarray}
The reason why we can actually go this step lies on the weak binding energy
of a single exciton. It was only one third of the depth of the potential 
for the first bound exciton, and will be even smaller as more excitons are tied.   

For each number $K$ of excitons, 
this equation must be solved self-consistently due to the dependence of $\mu
_K$ on $\phi _K$. Such a task must be performed numerically. 
In this way we have obtained both $\phi _K$ and the energy $\varepsilon _K$ up
to $K=4$. As expected, we obtain a set of metrics, $|\mu _1 (\phi _1)|^2$, ...
, $|\mu _4  (\phi _4)|^2$ which spread their inhomogeneous core
for increasing $K$. This implies
potentials $V_1(\phi _1)$, ... , $V_4(\phi _4)$ which are successively 
less attractive and more extended as shown in Fig. \ref{fig1}.  
As a direct consequence, the states 
$\mu _1\phi_1$, ... , $\mu _4\phi_4$ are progressively less localized.   

Once we have the self-consistent solution for an exciton in the presence of
both the flux quantum and a background of $K-1$ excitons, the mean field
wave function describing $K$ excitons bound to the defect is just 
\begin{eqnarray}
\Phi (z_1, ... ,z_K)= \phi _K(z_1) ... \phi _K(z_K)  
\end{eqnarray}
The energy $E_K$ of this state must be computed with a little care. If we 
just considered $E_K=K\varepsilon _K$, the exciton-exciton 
repulsion would be overestimated. Instead, we must pill up
the excitons one by one in the final state, summing up all the energies required
in the operation. We must then compute the expected values of 
effective Hamiltonians built up with $0,1,2,...,K-1$ excitons which are 
identical to the ones obtained self-consistently in the presence of a 
background of $K-1$ excitons. Labeling the corresponding effective Hamiltonians  
as ${\cal H}_1(\phi _K)$, ... , ${\cal H}_K(\phi _K)$, the energy becomes  
\begin{eqnarray}
E_K=\sum _{j=1} ^K \langle \phi _K|{\cal H}_j(\phi _K)|\phi _K \rangle .
\label{energy}
\end{eqnarray}
Table \ref{tableI} gives our results for each term of Eq. (\ref{energy}) as
well as for the energies $E_K$. Since ${\cal H}_K(\phi _K)\equiv H_K$, 
the diagonal of the left side of the table directly gives the eigenvalues
$\varepsilon _K$. 
Our results verify all the expected behaviors about progressively less 
bound excitons and decreasing total energy as $K$ increases. Repulsion 
between different excitons can also be drawn from Table \ref{tableI}. 

As discussed above, by repeating the procedure indefinitely, for $K =
\infty$, one should obtain the classical skyrmion of the NL$\sigma $M. This
is obviously impossible from the practical point of view but we can use 
a Wynn's algorithm \cite{betsuyaku} to extrapolate our results in Table
\ref{tableI} and estimate 
their asymptotic limit for the texture with $K \rightarrow \infty $. 
Such extrapolation (also included in table \ref{tableI}) gives 
$E_\infty =-1.03 \times 4 \pi \rho _s$, while the NL$\sigma $M
gives for skyrmions an energy $-4 \pi \rho _s$ with respect to $|^0F\rangle $.
Taking into account the intrinsic uncertainties of the extrapolation
procedure, this result can be considered as satisfactory enough to conclude that
our scattering procedure recovers the NL$\sigma $M skyrmion when infinite
excitons are bound to the defect.

The energies displayed in Table \ref{tableI} 
are significantly lower than the ones obtained from
Hartree-Fock calculations\cite{Fertig} or variational
procedures.\cite{Abolfath} Therefore, we would obtain larger textures
than those predicted by previous calculations. 
By extrapolating the energies given in Table
\ref{tableI}, we get, for the $g$-factor of GaAs, a skyrmion with $K=7$
for $B=4$T and $K=5$ for $B=20$T. This is too large compared with the size
experimentally estimated\cite{Barret,Schmeller} because we are not
taking into account both the finite width of the wells and the effect of 
higher Landau levels. Both effects tend to reduce the strength of the
interactions with respect to the Zeeman term. This implies a smaller
size of the skyrmion what means a better agreement with experiments.  

We still owe an explanation of why the angular momentum of the excitons 
in their ground state equals the winding number of the spin textures resulting
from a condensation of many of those excitons. Hitherto, we
have built many-body wave functions
\begin{eqnarray}
\left[ O^\dagger _{\phi} \right]^K|^0 F\rangle =  
\left[\int dz \; f(|z|) e^{-i\theta} \Psi ^\dagger _\downarrow (z)
\Psi _\uparrow (z)  \right]^K |^0 F\rangle
\label{fluct}
\end{eqnarray}
containing a well defined number $K$ of excitons (i.e. of spin flips)
to preserve the symmetry under rotations in spin space 
around the direction of the magnetic field. 
The expected value of the in-plane (perpendicular to the magnetic field)
magnetization in these states is then zero everywhere. Nevertheless, 
for vanishing Zeeman coupling, when many electrons
find it cheaper to flip their spin, we have just seen that the energy 
required for the texture 
to tie or drop a single exciton ($\lim _{K \rightarrow \infty} 
E_K - E_{K \pm 1}$) vanishes. Then, a coherent
superposition of states containing whatever number of excitons is allowed, 
and the system develops order in the magnetization. This coherent state
is described by the BCS-like wave function \cite{Fertig},  
\begin{eqnarray}
\prod_{m \ge 0}  (c^+_{m+1,\uparrow} + u_m 
e^{i \varphi} c^+_{m, \downarrow})\: | 0 \rangle .
\label{cohert}
\end{eqnarray}    
The projections of this wave function onto the subspaces of states with 
well defined number $K$ of spin flips (excitons) are just our 
mean field states \cite{Oaknin} 
\begin{eqnarray}
\left[ O^\dagger _{\phi} \right]^K|^0 F\rangle =  
C \int_0^{2\pi}
d\varphi \; e^{-i K \varphi} \prod_{m \ge 0} (c^+_{m+1,\uparrow} + u_m 
e^{i \varphi} c^+_{m , \downarrow})\: | 0 \rangle .
\end{eqnarray}    
$c^+_{m,\sigma}$ is the representation of
the electron field operator $\Psi ^\dagger_{\sigma} (z)$ in the 
states with well defined angular momentum, $\varphi_m$, given by 
Eq. (\ref{varphi}). $|0\rangle$ is the vacuum state, and $C$ is a constant. 
The degrees of freedom $f(|z|)$ in (\ref{fluct}) and $u_m$ in (\ref{cohert}) 
are simply related to each other by the change of representation. 
Note that excitons with unit angular momentum, only mix electron
states differing also in one unit of angular momentum. 

The BCS-like wave function (\ref{cohert}) describes, for slowly decaying 
$f(|z|)$, a spin texture with topological charge $1$. The decay 
$f(|z|) = 1/|z|$, that we have found in the previous section 
to minimize the energy for vanishing Zeeman coupling, is an antiskyrmion.  
In this latest case, the BCS-like wave function (\ref{cohert}) reads simply 
in first quantization \cite{Oaknin} 
\begin{eqnarray}
\prod_{i=1} ^N  \left( \begin{array}{c} z_i \\ \xi e^{i \varphi}
\end{array} \right) |F \rangle 
\end{eqnarray}    
that is the usual representation for antiskyrmions. The parenthesis is the
spinor of the $i$-th electron and $z_i$ its position in the plane.
The parameter $\xi$ gives the size of the antiskyrmion. 

\section{Conclusions}

We have presented a theoretical framework to describe skyrmions in QHF 
as a condensate of spin-waves bound to a bare charged defect. The scheme can be 
easily applied to QHF corresponding to filling factors $1/(2p+1)$, 
with $p$ integer.   

The low lying spectrum of neutral excitations of a QHF (spin-waves)   
has been mapped to that of free moving spin excitons of vanishing electrical
dipolar moment. We have then studied their scattering by a bare 
quasiparticle, and we have found that it may bind them. So the
charged excitation carries a spin texture. The bound excitons interact 
with the bare defect of charge as well as with each other. 
The effects of the former are: 

i) Pauli principle does not allow the excitons to be at the same position 
as the defect. We describe this non-uniform space where the exciton  
moves by an inhomogeneous metric 
in the Hilbert space of its wave functions.  

ii) The defect of charge leaves a quantum of magnetic flux unbalanced. 
This introduces an Aharonov-Bohm phase that gives a 
unit of angular momentum to the excitons in their ground state. 
The resulting texture has winding number equal to 1. 

iii) The localization of the spin of the bare quasiparticle implies a 
waste of exchange energy. The exchange energy that may 
be gained by a smoother alignment of 
spins is felt as an attractive potential where the excitons become bound. 
This potential preserves the symmetry under rotations in spin space in the
inhomogeneous space described in i). 

In this work we have deduced explicit expressions for both the 
inhomogeneous metric and the potential. Our framework is supported by comparison 
with results obtained by numerical calculations. 

To take into account the interaction among bound excitons we have used a mean field
approximation for them. We find that it works quite well not only for textures
containing many excitons, but also for those others containing a few of them. 

As a final comment, we must stress that all the states we obtain have a
well defined third component of the total spin. 
Therefore, 
it is not possible to associate to those states a local vector field having the 
characteristics of a spin texture. In order to get it, it is necessary to
allow for linear combinations of states with different values of the third
component of the spin.\cite{Oaknin} This is possible for 
vanishing Zeeman energies that favors states containing many excitons,  
because all of them are almost degenerate in energy. The resulting texture
is a skyrmion. 

\section{Acknowledgements}
Work supported in part by MEC of Spain under contract No. PB96-0085. 
J.H.O. is grateful to D. Orgad for helpful discussions.

\appendix
\section{ Coefficients $\alpha _m^{(K)}$ for the variation of the metric
}

The coefficients $\alpha _m^{(K)}$ giving the variation of the metric in 
Eq.(\ref{Deltamu}) are rational functions  of the variables 
\begin{eqnarray}
{\beta} _{m'}^{(K)}=\int dz|\phi _{K} (z)|^2 |\varphi _{m'} (z)|^2
\end{eqnarray}
with $m'$ going from 1 to infinity. 
Up to 4 excitons, these rational functions are:
\begin{eqnarray}
\alpha _m^{(1)}& =& 0 \\
\alpha _m^{(2)}& =& \frac{\beta _m^{(2)}}{\sum_{m'>0} \beta _{m'}^{(2)}}  \\
\alpha _m^{(3)}& =& \frac{\beta _m^{(3)} \sum_{m'} 
\beta _{m'}^{(3)}- [ \beta _{m}^{(3)}]^2}{\left(\sum_{m'}\beta _{m'}^{(3)}
\right)^2-\sum_{m'}[\beta _{m'}^{(3)}]^2} \\
\alpha _m^{(4)}& =& \frac{\beta _m^{(4)} \left( \sum_{m'} 
\beta _{m'}^{(4)} \right) ^2- 2[ \beta _m^{(4)}]^2 
\sum_{m'} \beta _{m'}^{(4)} + 2\sum_{m'}
[ \beta _{m'}^{(4)}]^3 - \beta _m^{(4)}  
\sum_{m'} [ \beta _{m'}^{(4)}]^2 }{
\left( \sum_{m'}\beta _{m'}^{(4)}\right)^3-
3\left( \sum_{m'}\beta _{m'}^{(4)}\right) \sum_{m'}[\beta _{m'}^{(4)}]^2
+2\sum_{m'}[\beta _{m'}^{(4)}]^3 } 
\end{eqnarray}

\begin{figure}
\caption{ Effective potential $\nabla ^2\mu _K (z)/\mu _K (z)$ for
$K=1,..,4$ as a function of the distance $|z|$ 
(in units of the magnetic length)
to the charged defect. The eigenvalues $\varepsilon _K$ are also shown.
}
\label{fig1}
\end{figure}

\begin{table}
\caption{ Binding energies $\langle \phi _K|{\cal H}_j(\phi _K)|\phi _K \rangle $
of the $j$-th exciton in the $K$-th step (see text) and total energy  $E_K$
of the texture with $K$ excitons bound to the charged defect in units of 
$4 \pi \rho _s$. The total energy $E_{K \rightarrow \infty}$ obtained by an
extrapolation (see text) is also included.
The diagonal of the left side gives the eigenvalues $\varepsilon _K$. }
\begin{tabular}{|c|cccc|c|} 
& $\langle \phi _K|{\cal H}_1(\phi _K)|\phi _K \rangle $ 
& $\langle \phi _K|{\cal H}_2(\phi _K)|\phi _K \rangle $ 
& $\langle \phi _K|{\cal H}_3(\phi _K)|\phi _K \rangle $ 
& $\langle \phi _K|{\cal H}_4(\phi _K)|\phi _K \rangle $ 
& $E_K$ 
\\ \hline
K=1 & -0.311 & &  & & -0.311 \\
K=2 & -0.304 & -0.191 & & & -0.495 \\
K=3 & -0.295 & -0.201 & -0.138 & & -0.634 \\
K=4 & -0.287 & -0.205 & -0.149 & -0.109 & -0.750 \\
\hline
K$\rightarrow \infty $ & & & & & -1.030   

\end{tabular}
\label{tableI}
\end{table}

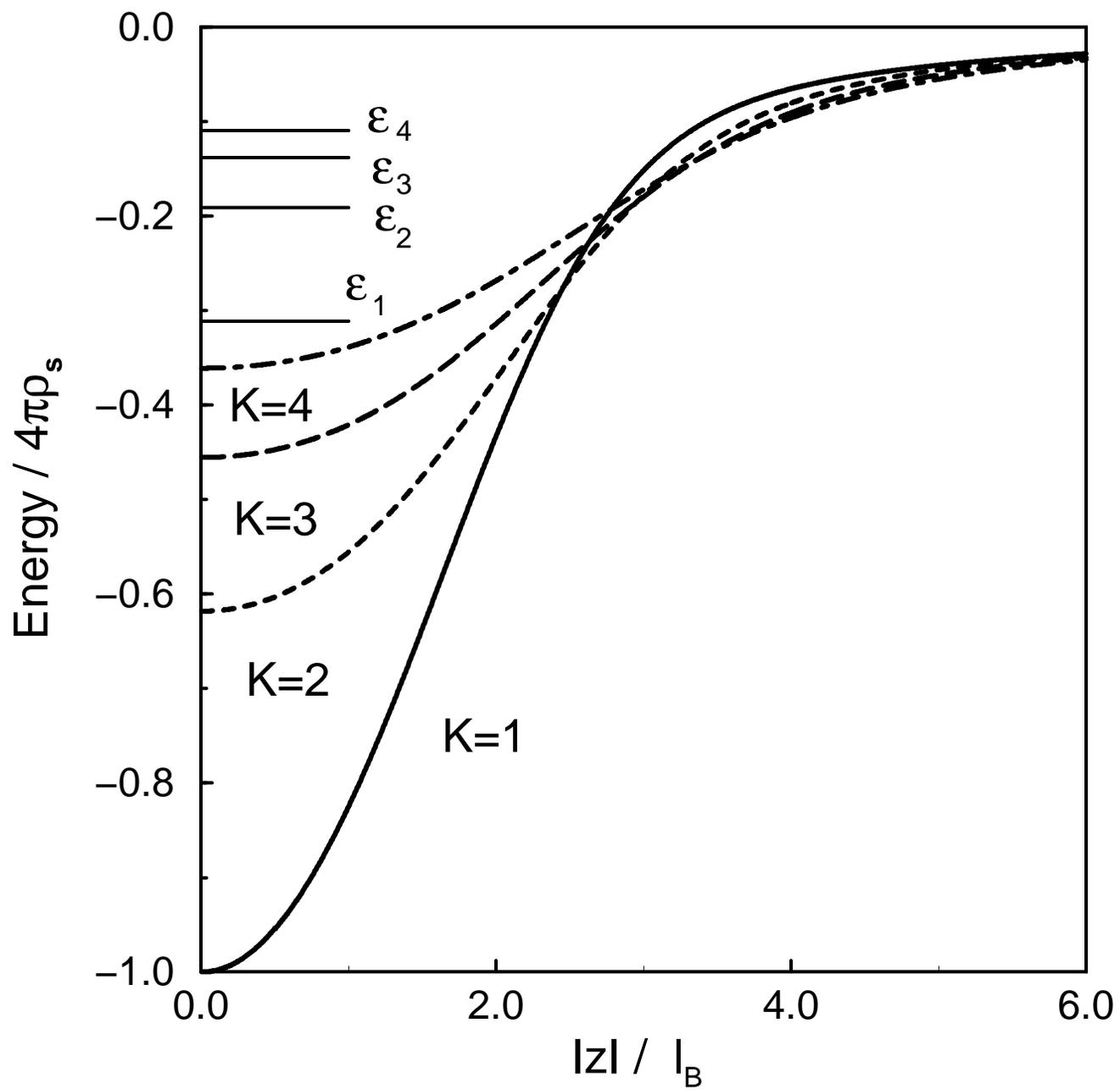
\end{document}